\documentclass[9pt]{pasa}%
\usepackage{graphicx,subfig}

\usepackage[authoryear]{natbib}
\usepackage{graphicx}

\jid{PASA}
\doi{10.1017/pas.\the\year.xxx}
\jyear{\the\year}

\usepackage{hyperref}
\hypersetup{colorlinks,citecolor=blue,linkcolor=blue,urlcolor=blue}

\title[The Origin of Isolated Millisecond Pulsars]{On Isolated Millisecond Pulsars Formed by the Coalescence of Neutron Stars and Massive White Dwarfs}

\author[Shengnan Sun et al.]{Shengnan Sun$^{1}$\thanks{Email: 2438754411@qq.com},Lin Li$^{1, 2, 3}$, Helei Liu$^{1}$, Guoliang L\"{u}$^{1}$,Zhaojun Wang$^{1}$,Chunhua Zhu$^{1}$\thanks{Corresponding author: Chunhua Zhu, Email: chunhuazhu@sina.cn} \\
\affil{$^1$School of Physical Science and Technology, Xinjiang University, Urumqi, 830046, China}
\affil{$^2$Xinjiang Astronomical Observatory, Chinese Academy of Sciences, 150 Science 1-street, Urumqi, Xinjiang 830011, China}
\affil{$^3$Key Laboratory of Electronic Equipment Structure Design, Ministry of Education, Xidian University, Xi¡¯an, Shaanxi, 710071}
}%

\begin{document}

\begin{abstract}
This paper uses population synthesis to investigate the possible origin of isolated millisecond pulsars (IMSPs) as born from the coalescence of a neutron star (NS) and a white dwarf (WD). Results show that the galactic birth - rate of IMSPs is likely to lie between $5.8\times10^{-5}$ $\rm yr^{-1}$ and $2.0\times10^{-4}$ $\rm yr^{-1}$, depending on critical variables, such as the stability of mass transfer via the Roche lobe and the value of kick velocity. In addition to this, this paper estimates that the solar mass of IMSPs can range from $1.5$ and 2.0 $M_{\odot}$, making them more massive than other 'normal' pulsars. Finally, the majority of IMSPs in our simulations have spin periods ranging from several to 20 milliseconds, which is consistent with previous observations.

\end{abstract}


\maketitle

\section{Introduction}
In 2017, GW170817, a gravitational wave produced by the merging of two neutron stars (NS), was observed by Advanced LIGO/Virgo \citep{2017PhRvL.119p1101A}. Its transient counterparts were detected across almost the entire electromagnetic spectra \citep[e.g.,][]{2017ApJ...848L..13A,2017ApJ...848L..33A,2017ApJ...848L..17C}, hence hailing the advent of a new era in multi-messenger astronomy.

Notably, only two seconds after the discovery of GW170817, a short $\gamma$-ray burst (GRB) was detected by both the Fermi and INTEGRAL space telescopes \citep{2017ApJ...848L..13A}. Traditionally, GRBs are divided into long GRBs and short GRBs. Long GRBs ($\rm T_{90} \geq2 s$) result from the death of massive stars and their accompanying supernovae. Short GRBs, however, ($\rm T_{90} \leq2 s$) are a result of the merging of two compact objects \citep{1986ApJ...308L..43P,1989Natur.340..126E,1992ApJ...389L..45P}. Therefore, the discovery of a short GRB provides the tantalizing prospect that GW170817 may have originated from a double NS merger. According to \cite{2017PhRvL.119p1101A}, the product of GW170817 was a compact object with a solar mass of 2.7 $M_\odot$. It is possible that this object may form a millisecond pulsar (MSP), a magnetar, in the future, or it may even rapidly collapse into a black hole \citep{1998PhRvL..81.4301D,2006Sci...311.1127D}, but due to its sheer distance, there are, no telescopes or astronomical equipment in the world capable of confirming its current state.

In 2006, the works of \cite{2006Natur.444.1044G}, \cite{2006Natur.444.1050D} and \cite{2006Natur.444.1053G} published the discovery of a special $\gamma$-ray burst: GRB060614. It was a long GRB ($\sim100\rm s$) and was not associated with any supernova \citep{2006Natur.444.1044G,2006Natur.444.1050D,2006Natur.444.1053G}. About this, \cite{2007MNRAS.374L..34K} suggests that GRB060614 may have originated from the merging of a NS and a massive white dwarf (WD). Currently, a total of 10 GRBs have been observed as having GRB060614-esque properties \citep{2016ApJ...832..136R}.

What's more, the subset of NS+WD binaries, resulting from a merger, would be likely source for so-called 'calcium rich gap' transients. These are a class of optical events characterised by sub luminous type-I supernovae, such as SN 2005E \citep{2010Natur.465..322P,2012ApJ...755..161K}.

In theory, merging a NS and a massive WD could produce an isolated millisecond pulsar (IMSP) \citep{1984A&A...139L..16V}. IMSPs are a special type of pulsar. According to data from \citep{2005AJ....129.1993M}, approximately 2600 different pulsars have been observed, most of which are 'normal pulsars' with pulse periods of P $\sim$ 0.1s to 10s. However, there are roughly 325 other pulsars, which exhibit pulse periods of $\sim$1.4 to 20 ms \citep{2013Ap&SS.346..119P}. These are so-called MSPs. The typical age ($\tau_{\rm c}$) of 'normal pulsars' is $10^{7}$ yr, with a surface magnetic field strength (B) of $10^{12}$ G. For MSPs, however, these values are $10^{9}$ yr and $10^{8}$ G, respectively \citep{2008LRR....11....8L}. Generally, MSPs are thought to be NSs with a high rotation rate, being formed from the accreting matter of a NS and its companion star \citep[See:][]{1982Natur.300..728A,1991PhR...203....1B}.

For this reason, one might surmise that all MSPs evolved from binary systems. However, according to the ATNF (Australia Telescope National Facility) catalogue, approximately 1/3 of MSPs are isolated \citep{2005AJ....129.1993M}.  Their origins are highly debated, with different scholars proposing a variety of different possible scenarios.

Firstly, \cite{1984A&A...139L..16V} proposed that IMSPs were a product of gravitational wave emissions after the merging of a NS and a massive WD. However, after the millisecond pulsar
PSR 1957+20 was discovered by \cite{1988Natur.333..237F}, it was proposed that pulsar wind was the major cause of IMSP formation \citep{1988Natur.334..225K}. In other words, a low-mass helium WD (with a solar mass $\sim0.02\rm M_\odot$) would be left when the mass-transfer results in a low-mass X-ray binary. Due to the high energetic radiation emitted from the MSP, the WD would then be completely ablated. This process was known as the Standard Model. However, \cite{1998ApJ...499L.183S}  found that the timescale for ablation provided in the Standard Model was too long for a NS to evolve into an IMSP in Hubble time \citep[See also][]{2013ApJ...775...27C}. Though the $\sim9$ hour orbit of PSR 1957+20 has a relatively short ablation timescale of merely $3\times10^{7}\rm yr$ \citep{1991ApJ...380..557R}, it is unlikely that this could truly be observed in such a short time phase.

Secondly, \cite{1991PhR...203....1B} proposed another possible evolutionary scenario. They believed that the MSPs formed from high-mass X-ray binaries can evolve into IMSPs when such binaries are disrupted by a core-collapse supernova from their companion stars. Although this scenario may be suitable for some IMSPs in these systems, it is unlikely to represent the most common formation models \citep{2010MNRAS.407.1245B}. Recently, a possible solution in the form of a triple-star formation model was discussed by \cite{2011MNRAS.412.2763F}, based on ideas previously put forward by \cite{1986MNRAS.220P..13E}. In their model, the orbit of a triple star system expands when materials transfer from the donor star to the NS. Subsequently, the accreting NS will then evolve into a MSP. If the triple star system becomes dynamically unstable, then it is possible for the MSP to be ejected, hence forming an IMSP. \cite{2011ApJ...734...55P}, however, contradict this model, stating that these phenomena in fact contribute very little to the formation of IMSPs.

As can be seen from the discussion above, none of the three scenarios above have been able to provide a definitive explanation for the origin of IMSPs. In the first model, \cite{2001A&A...375..890N} estimate that there are, theoretically, $\sim2.2\times10^{6}$ NS+WD binaries in the Galaxy and the merger rate of these systems ranges from about $1.0\times10^{-6}\rm yr^{-1}$ \citep{2004MNRAS.354...25C} to 1.4$\times10^{-4}\rm yr^{-1}$ \citep{2001A&A...375..890N}. Similarly, \cite{2009arXiv0912.0009T} states, with 95\% confidence, that the lowest galactic merger rate for NS+WD systems is 2.5 $\times10^{-5}\rm yr^{-1}$.

This paper builds on the previous research mentioned above, undertaking its own investigation into the possibility of forming IMSPs via the merging of NSs and WDs. Section 2 below presents both the authors¡¯ assumptions and details on the modelling algorithms. This is then followed by a set of results in \S3 and final conclusions in \S4.

\section{Models}

In this paper, the authors draw on the rapid binary star evolution (BSE) code, as expounded in \cite{2000MNRAS.315..543H}, \cite{2002MNRAS.329..897H}, and updated by \cite{2006MNRAS.369.1152K}. Unless specifically mentioned, our default input parameters are also based on those found in the above literature.

\begin{table}
\begin{center}
\setlength{\abovecaptionskip}{0pt}
\setlength{\belowcaptionskip}{10pt}
\caption{Parameters of the population models for NS+WD binaries.}
\begin{tabular}{ccccc}\hline\hline
Case            &          $q_{\rm c}$\ or\ $M^{\rm c}_{\rm WD}$       &  $\rm\sigma_{k}(kms^{-1})$          &        \\ \hline
case 1&$q_{\rm c}=0.628$ & 190 \\
case 2&$M^{\rm c}_{\rm WD}=0.37$    &190 \\
case 3&$M^{\rm c}_{\rm WD}=0.2$  &190 \\
case 4&$q_{\rm c}=0.628$  &265 \\
case 5&$M^{\rm c}_{\rm WD}=0.37$ &265\\
case 6&$M^{\rm c}_{\rm WD}=0.2$  &265 \\

\hline\hline
\end{tabular}
\label{tab:results}
\end{center}
\end{table}

\subsection{Kick Velocity}

During its formation, non-spherical symmetry of a NS creates additional 'kick' velocity. The physical origin of this non-spherical symmetry, however, remains somewhat enigmatic. In 1975, \cite{1975Natur.253..698K} suggested that these kicks may have a dichotomous nature, which was later confirmed by \cite{1997A&A...322..477H} and \cite{2002ApJ...573..283P}. Despite this, kick velocity cannot be easily controlled during observation, due to its numerous and complicated selection effects.

Typically, the distribution of the kick velocity is a Maxwellian with a dispersion ($\sigma_{\rm k}$) of

\begin{equation}\label{}
P(\nu_{\rm k})=\sqrt{\frac{2}{\rm \pi}}\frac{\nu_{\rm k}^{2}}{\sigma_{\rm k}^{3}}e^{-\nu_{\rm k}^{2}/2\sigma_{\rm k}^{2}}.
\end{equation}
In \cite{1997MNRAS.291..569H}, analysis of the proper motion of approximately 100 pulsars found $\sigma_{\rm k}$ to be equal to 190 $\rm kms^{-1}$. \cite{2005MNRAS.360..974H} however, examined the proper motion of 233 pulsars and found that the kick velocity could be described by a single Maxwellian with $\sigma_{\rm k}=265\rm km s^{-1}$. In the present investigation, we apply different velocity dispersions in different cases.

\subsection{Evolution of NS+WD Binary Systems}

NSs and WDs are both stellar remnants. Once binary systems are formed from them, there can often be disastrous consequences. In NS+WD systems, gravitational wave radiation can cause the orbital angular momentum ($J_{\rm orb}$) to decay. \cite{1971ApJ...170L..99F} noted that the decay ratio of $J_{\rm orb}$ was the following formula, where $\rm c$ is the speed of light and $a$ is the separation of the binary system.

\begin{equation}
  \frac{\dot{J_{\rm GB}}}{J_{\rm orb}}=-\frac{32 \rm G^{3}}{5\rm c^{5}}\frac{M_{\rm NS}M_{\rm WD}M}{a^{4}}
  \label{gr}
\end{equation}
As the orbital period shrinks, the WD fills its Roche lobe and begins to act as a donor, transferring its mass to the NS. This mass transfer can be either dynamically stable or unstable -- at this point, either outcome is possible. Based on an investigation into stable mass transfer, using polytropic models, \cite{1987ApJ...318..794H} and \cite{2002MNRAS.329..897H} concluded that if the mass ratio of the components ($q = M_{\rm donor}/M_{\rm gainer}$)  is larger than a certain value, $q_{\rm c}$, at the onset of Roche lobe overflow, then the mass transfer will be dynamically unstable. Otherwise, the mass transfer will be stable. According to \cite{2002MNRAS.329..897H}, the value of $q_{\rm c}$, in a NS+WD binary system is 0.628. However, based on the isotropic re-emission mechanism\footnote{By assuming that the accreting limit of the gainer precisely equals Eddington limit, the additional transferred matter is unbound to the binary system and its gravitational energy is released \citep{1997A&A...327..620S,1999A&A...350..928T}.}, \cite{2012A&A...537A.104V} put forward even stricter criteria for mass transfers in a NS+WD system to be stable, i.e. the critical mass of the WD ($M^{\rm c}_{\rm WD}$) donor must be 0.37 $M_\odot$. Recent scholarship into the angular momentum of material lost in disc winds, such as \cite{2017MNRAS.467.3556B}, has found that the $M^{\rm c}_{\rm WD}$ may also be equal to 0.2 $M_\odot$, significantly lower than previously thought. If the mass transfer does, for some reason, become dynamically unstable, then the WD will be tidally disrupted by the NS, which, in turn, may lead to a merger and a gravitational wave event \citep{2009PhRvD..80b4006P,1984A&A...139L..16V}. However, according to \cite{2016MNRAS.461.1154M}, in most circumstances, the mass accreted by a NS is insufficient to induce a gravitational collapse. It is often capable, however, of simply increasing the rotational velocity of the NS by several milliseconds. For this reason, it is critically important to consider the parameters of dynamically unstable mass transfer in our investigative models.

In order to discuss the effect of $q_{\rm c}$ on the formation of IMSPs, the authors calculated a total six possible cases, each of which contain different combinations of mass and kick velocity values, such as those mentioned above (See Table \ref{tab:results}).

\subsection{Post-Merger NS}

As mentioned above, this paper is built on the assumption that the merging of a NS and a WD can produce an IMSP. After a merge, \cite{2012MNRAS.419..827M} estimated that about 20-50\% of the WD's matter is accreted by the NS and then the remainder is ejected as energy, released by both gravity and nuclear reactions during the tidal disruption of the WD. Here, we assume that the post-merger mass of the NS is equal to $M_{\rm NS}$ + 0.5$M_{\rm WD}$.

After merging, the spin ($P_{\rm s}$) of the NS is also affected by its surrounding matter. There is a great deal of scholarship on the interaction between rotating magnetized NSs and their surrounding matter \cite[e. g.,][]{1972A&A....21....1P,1975A&A....39..185I, 1992ans..book.....L,1999ApJ...514..368L}. In these works, the value of $P_{\rm s}$ chiefly depends on both the NS' mass accretion rate and  also its magnetic field. What's more, during the merging process, the mass accretion rate ( $\sim 10^{-2}M_\odot \rm yr^{-1}$) is significantly higher than the Eddington Accretion Rate \citep{2012MNRAS.419..827M}. In 1982, \cite{1982SvA....26...54L} investigated magnetized NSs with a super-Eddington rate of accretion, giving the NS' equilibrium spin period as

 \begin{equation}\label{}
  P_{\rm s}^{\rm eq}=1.76\times10^{-1}\mu^{2/3}_{30}M_{\rm NS}^{-2/3}\rm s
\end{equation}

In the above, $\mu_{30}=\mu/(10^{30}\rm Gcm^{3})$, where $\mu=B_{\rm NS}R_{\rm NS}^{3}/2$ denotes the magnetic dipole momentum, $B_{\rm NS}$ is the magnetic field and $R_{\rm NS}$ is the radius of the NS. In this particular case, the $R_{\rm NS}=10^6$ cm. After the merger, the spin then decreases due to a brake in the current \citep{1993ConPh..34..131B}. Hence, the evolution of spin can be given approximately as

\begin{equation}\label{}
 \frac{{\rm d}P_{\rm s}}{{\rm d}t}=\frac{10^{-39}B_{\rm NS}^{2}}{P_{\rm s}}.
\end{equation}

The evolution of NSs' magnetic fields is still unknown. Using their death and spin-increase line as evidence, \cite{1997MNRAS.284..741U} calculated a decrease in a NS' magnetic field strength as it evolved into a MSP. Their calculation was as follows:

\begin{equation}\label{}
   B(\rm t)=\emph B_{0}(\frac{\emph t_{0}}{\emph t_{0}+\emph t})^{\beta}
 \end{equation}

In the above, $ B_{0}, t_{0}$ and $t$ respectively denote the initial field strength, the time-scale of decay and the age of the NS. $\beta$ is a free parameter. Based on figures from the ATNF, pulsars which age less than $10^5$ yr have magnetic fields of between roughly $10^{12}$ and $10^{13}$ G, averaging out at approximately $8\times 10^{12}$ G. Hence, this work takes the value of $B_0$ as $8\times10^{12}$ G and $t_0$ as $10^5$ yr. In the majority of cases, the time it takes for a NS to form and then merge with its companion star may range from around $10^6$ to $10^8\rm yr$. For this reason, the authors follow in the footsteps of \cite{1997MNRAS.284..741U},  taking $\beta$ as 1.0.  Whilst merging, the magnetic field of the NS may either increase via a so-called 'winding-up' process \citep{2016MNRAS.462L.121O}, or decrease due to an enhanced Ohmic dissipation of accreted matter. However, to the authors' knowledge, there are currently no models available to simulate this event. For this reason, the models used in this paper assume that magnetic fields do not change during the merging process.

\section{Results}

In order to understand more about the birth-rate of IMSPs and their physical properties, the authors use population synthesis to simulate the evolution of $10^7$ binary systems.

The cases considered in the present study show a similarity to those in \cite{2006MNRAS.372.1389L,2008ApJ...683..990L,2009MNRAS.396.1086L,2012MNRAS.424.2265L,2013ApJ...768..193L}. Notably, the authors use a simple approximation to the initial mass function (IMF), based on \cite{1979ApJS...41..513M}. The primary mass is then generated using the formula suggested by \cite{1989ApJ...347..998E}.  The distribution of separations is given by $\rm log \emph a=5\emph X + 1$, where $X$  is a random variable between 0 and 1 and $a$ is orbital separation in units of $R_{\odot}$.  In our models, all binaries have initial circular orbits and the metallicity $Z$ is set to 0.02 for PopulationI stars.

Furthermore, in the case of a constant star formation rate, the authors assume that one binary, with a primary body more massive than 0.8$\rm M_{\odot}$, is formed annually in the Galaxy (\cite{1993ApJ...418..794Y}; \cite{1995MNRAS.272..800H}; \cite{2016JApA...37...22Y}).

\subsection{NS + WD Binaries}

There are three possible evolutionary pathways for NS+WD binaries. Firstly, the WD is unable to fill its Roche lobe, due to its long orbital period within Hubble time. Secondly, the WD can fill its Roche lobe and its matter can undergo a stable transfer to the NS, hence evolving into an ultra-compact X-ray binary. Thirdly, the WD successfully fills its Roche lobe and the merging process begins. It is the latter of the three pathways that is the focus of this investigation.

According to ATNF, there are a total of 147 binaries, composed of one pulsar and one WD \citep{2005AJ....129.1993M}. Figure \ref{fig:m2po} below gives a comparison of WD masses and orbital periods between the NS+WD binaries in our simulation and those from ATNF. On average, our results tend to show overall larger masses than those observed in other systems, where figures are usually recorded as around $0.2$ to $0.3\rm M_{\odot}$. For example, there were even systems in which the mass of the WD is between $0.4$ and $0.8\rm M_{\odot}$. Since these WDs have smaller radii, they will never fill their Roche-lobes in Hubble time. In fact, the majority of NSs recorded in these systems appear to have already passed their deadline, hence should no longer be observed as pulsars.

If the WDs in NS + WD binaries fill their Roche lobes within Hubble time, then mass transfer shall inevitably occur. Once mass transfer begins, if $q<q_{\rm c}$ or $M_{\rm WD}<M^{\rm c}_{\rm WD}$, then it will be dynamically stable and the original binaries will, in turn, evolve into new ultra-compact X-ray binaries. At present, there are a total of 30 known ultra-compact X-ray binaries and formation candidates in the Galaxy \citep{2007A&A...469..807L,2010NewAR..54...87N}. The evolution of such ultra-compact X-ray binaries has been investigated by previous researchers, such as \cite{2008AstL...34..620Y,2012A&A...537A.104V,2017ApJ...847...62L}. Contrastingly, if $q>q_{\rm c}$ or $M_{\rm WD}>M^{\rm c}_{\rm WD}$, then a dynamical mass transfer will occur stably and the NS and WD will begin to merge into an IMSP.

\begin{figure*}[htp]
\begin{center}
\includegraphics[totalheight=3.5in,width=3.6in,angle=-90]{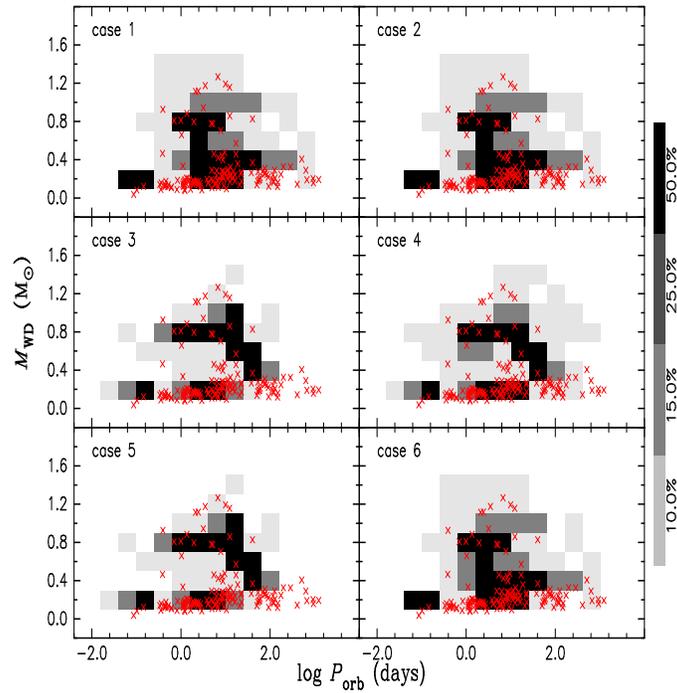}
\caption{Grey-scale maps of WD masses and orbital periods in NS+WD binaries. The red multiplication signs $(\times)$ represent observational values for the binaries composed of a pulsar and a WD. The observational data comes from the ATNF, as cited in \cite{2005AJ....129.1993M}.}
\label{fig:m2po}
\end{center}
\end{figure*}

\subsection{IMSP Population}
\begin{figure*}[htp]
\begin{center}
\includegraphics[totalheight=3.5in,width=3.6in,angle=-90]{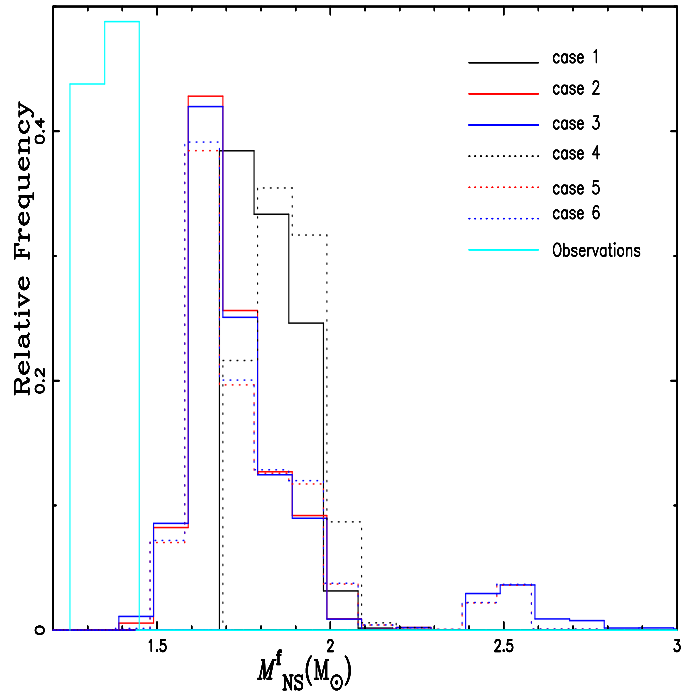}
\caption{The mass distribution of IMSPs in the paper's investigative models. The observational data consists of currently-known normal stars' pulsar masses, which come from https://stellarcollapse.org/nsmasses
\citep{2012AIPC.1484..319L}.}
\label{fig:mass}
\end{center}
\end{figure*}

In our investigative models, an IMSP's mass equals $M_{\rm NS}$ + $0.5M_{\rm WD}$. Figure \ref{fig:mass} uses a chart to illustrate the distribution of IMSP mass in our models. From the chart, one can see that when $q_{\rm c}=0.628$, the solar mass of the IMSP ranges from 1.7 to 2.0 $M_\odot$, whereas the mass of other models lies between 1.5 and 1.7 $M_\odot$. This shows that IMSPs are generally more massive than normal pulsars, which themselves have a total mass of around $1.4 M_\odot$. Unfortunately, academia has not yet been able to provide any accurate measurements for the exact mass of an IMSP. What's more, the highest left peak in Figure \ref{fig:mass} reaches approximately 2.5$\rm M_{\odot}$. This highlights the possibility that the progenitors of NSs may have had a higher initial mass and shorter initial period, causing NSs to gain a total solar mass of approximately $>2.0 \rm M_{\odot}$ during their pre-merging period.

\begin{figure*}[htp]
\begin{center}
\includegraphics[totalheight=3.5in,width=3.6in,angle=-90]{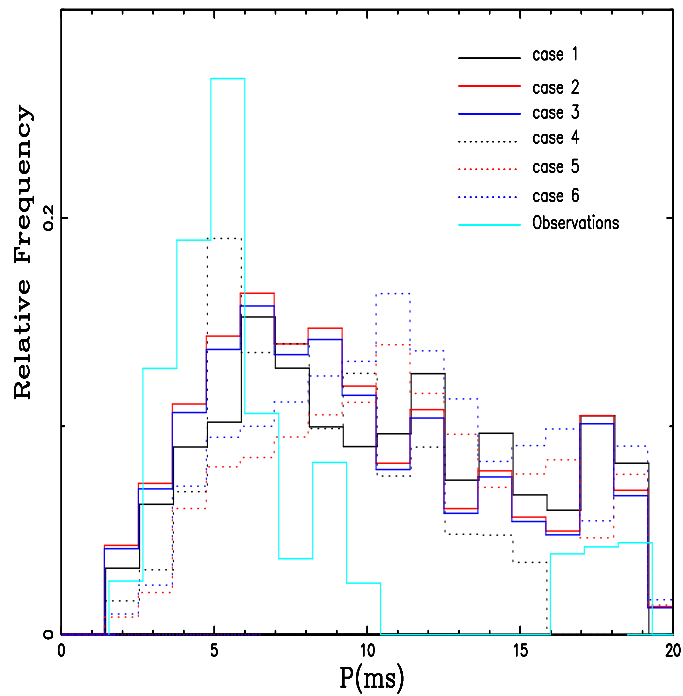}
\caption{The distributions of spin periods for IMSPs in different models. The data included comes from the ATNF, as cited in \cite{2005AJ....129.1993M}.}
\label{fig:spin}
\end{center}
\end{figure*}

In addition to mass, spin periods are also one of the most important physical parameters of an IMSP. Figure \ref{fig:spin} presents spin distributions for the IMSPs recorded in the present study. In general, the data correlates with previous observations, but some peaks in our model seem to show a somewhat slower spin than those observed in previous studies. This leads one to surmise that the widely accepted figures for the magnetic field of a post-merger NS could possibly be an overestimation. Nevertheless, the magnetic field of a NS is notoriously hard to determine. Furthermore, researchers have not yet reached a common understanding on the decay of an accreting NS' magnetic field. One possibility was proposed by \cite{1974AZh....51..373B},
who, through the examination of original magnetic fields, concluded that the decay of a NS may be caused by its accreted matter.

\subsection{Merger Rate}

Using the method of population synthesis, the present study estimates that the merger rate of NS+WD binaries lies between $5.8\times10^{-5}$ $\rm yr^{-1}$ (case 4) and $2.0\times10^{-4}$ $\rm yr^{-1}$(case 3). From case 1 to case 3, the numbers of NS+WD binaries also increased from 666 to 1981, thanks to differing critical values. In other words, one can conclude that the smaller the critical mass of a WD, the more IMSPs will be formed. In addition to this, comparing cases 1 with 4, 2 with 5 and 3 with 6 suggests that the higher the kick velocity is, the less IMSPs are formed. This is because the nascent NS' kick velocity plays a determining role in whether the binary can continue to survive or whether, if the kick velocity is too high, it will be disrupted by a supernova explosion.

Furthermore, the authors of the present survey also consider NS+WD mergers as a potential progenitor of calcium-rich supernovae. According to \cite{1998ARA&A..36..189K}, the Milky Way took $0-10 \rm M_{\odot}yr^{-1}$, to form. If one takes the median of this range, then that makes $5 \rm M_{\odot}yr^{-1}$ the rate of star formation. \cite{2012ApJ...755..161K}, however, put forward a possible a lower limit for calcium-rich gap events of $7\times10^{-7}\rm Mpc^{-3}yr^{-1}$.

To calculate the total star production rate, one can use the following formula from \cite{2004ApJ...613..200S}:

\begin{equation}\label{}
SFR=10^{9}a(t^{b}e^{-\frac{t}{c}}+de^{\frac{d(t-t_{0})}{c}})\rm M_{\odot}yr^{-1}Gpc^{-3}
\end{equation}
Therein, $t$ denotes the age of the Universe in Gyr and $t_{0}$ is the current age of the Universe. If one follows through with \cite{2004ApJ...613..200S}'s method and takes $t_{0}$ to be $13.47 \rm Gyr$, and the parameters $a=0.021, b=2.12, c=1.69$, and $d=0.207$, then one can infer that the rate of Ca-rich gap transients in the Galaxy is roughly $5\times10^{-4}\rm yr^{-1}$. This result is relatively close to our simulation-based estimations.

In general, our results are consistent with \cite{2001A&A...375..890N}. The merger rate of NS+WD binaries is approximately 3 to 10 times that of double NS systems \citep{1999MNRAS.309...26P}. Although gravitational waves are released during the merging of a NS and a WD, they can hardly be detected, due to their relatively low amplitude and frequency. It is their electro-magnetic counterparts that should be observed, as they eject significantly more observable matter. As for whether GRBs, such as GRB060614, originate from the merger of a NS and a WD binary, this hypothesis is certainly possible, but many more multiband observations are still needed until we get a more defined and clear-cut answer.

\section{Conclusion}

In this paper, the authors investigated the merging of neutron stars (NS) and white dwarfs (WD) as the possible origin of isolated millisecond pulsars (IMSP). This paper not only estimated that IMSPs' galactic birth-rate is between approximately $5.8\times10^{-5}$ $\rm yr^{-1}$ and $2.0\times10^{-4}$ $\rm yr^{-1}$, depending on variables such as stable mass transfer via the Roche lobe and kick velocity, but also predicted that the solar mass of IMSPs lies between $1.5$ and 2.0 $M_{\odot}$, which makes them more massive than normal pulsars. What's more, most of the IMSPs in our simulations have a spin period of several to 20 milliseconds, which is consistent with previous observations.

If it is true that the majority of IMSPs originate from the merging of NSs and WDs, then this event may trigger a GRB060614-esque reaction and produce gravitational waves. If it be possible to detect these in the future, then our field will be one step closer to a fuller understanding of gamma-ray bursts and gravitational waves.

\section*{Acknowledgements}
We would like to take this opportunity to thank an anonymous referee for all their useful comments.  This work received the generous support of the  National Natural Science Foundation of China, project No.11763007, 11473024, 11463005, 11863005
 and 11503008. We would also like to express our gratitude to the Tianshan Youth Project of Xinjiang No.2017Q014.


\begin{thebibliography}{l}
\scriptsize
\bibitem[Abbott et al.(2017)]{2017PhRvL.119p1101A} Abbott, B.~P., Abbott, R., Abbott, T.~D., et al.\ 2017, Physical Review Letters, 119, 161101
\bibitem[Abbott et al.(2017)]{2017ApJ...848L..13A} Abbott, B.~P., Abbott, R., Abbott, T.~D., et al.\ 2017, 848, L13
\bibitem[Alpar et al.(1982)]{1982Natur.300..728A} Alpar, M.~A., Cheng, A.~F., Ruderman, M.~A., \& Shaham, J.\ 1982, 300, 728
\bibitem[Arcavi et al.(2017)]{2017ApJ...848L..33A} Arcavi, I., McCully, C., Hosseinzadeh, G., et al.\ 2017 , 848, L33
\bibitem[Belczynski et al.(2010)]{2010MNRAS.407.1245B} Belczynski, K., Lorimer, D.~R., Ridley, J.~P., \& Curran, S.~J.\ 2010, 407, 1245
\bibitem[Beskin(1993)]{1993ConPh..34..131B} Beskin, V.~S.\ 1993, Contemporary Physics, 34, 131
\bibitem[Bhattacharya \& van den Heuvel(1991)]{1991PhR...203....1B} Bhattacharya, D., \& van den Heuvel, E.~P.~J.\ 1991, 203, 1
\bibitem[Bisnovatyi-Kogan \& Komberg(1974)]{1974AZh....51..373B} Bisnovatyi-Kogan, G.~S., \& Komberg, B.~V.\ 1974, 51, 373
\bibitem[Bobrick et al.(2017)]{2017MNRAS.467.3556B} Bobrick, A., Davies, M.~B., \& Church, R.~P.\ 2017, 467, 3556
\bibitem[Chen et al.(2013)]{2013ApJ...775...27C} Chen, H.-L., Chen, X., Tauris, T.~M., \& Han, Z.\ 2013, 775, 27
\bibitem[Cooray(2004)]{2004MNRAS.354...25C} Cooray, A.\ 2004, 354, 25
\bibitem[Cowperthwaite et al.(2017)]{2017ApJ...848L..17C} Cowperthwaite, P.~S., Berger, E., Villar, V.~A., et al.\ 2017, 848, L17
\bibitem[Dai \& Lu(1998)]{1998PhRvL..81.4301D} Dai, Z.~G., \& Lu, T.\ 1998, Physical Review Letters, 81, 4301
\bibitem[Dai et al.(2006)]{2006Sci...311.1127D} Dai, Z.~G., Wang, X.~Y., Wu, X.~F., \& Zhang, B.\ 2006, Science, 311, 1127
\bibitem[Della Valle et al.(2006)]{2006Natur.444.1050D} Della Valle, M., Chincarini, G., Panagia, N., et al.\ 2006, 444, 1050
\bibitem[Eggleton et al.(1989)]{1989ApJ...347..998E} Eggleton, P.~P., Fitchett, M.~J., \& Tout, C.~A.\ 1989, 347, 998
\bibitem[Eggleton \& Verbunt(1986)]{1986MNRAS.220P..13E} Eggleton, P.~P., \& Verbunt, F.\ 1986, 220, 13P
\bibitem[Eichler et al.(1989)]{1989Natur.340..126E} Eichler, D., Livio, M., Piran, T., \& Schramm, D.~N.\ 1989, 340, 126
\bibitem[Faulkner(1971)]{1971ApJ...170L..99F} Faulkner, J.\ 1971,170, L99
\bibitem[Freire et al.(2011)]{2011MNRAS.412.2763F} Freire, P.~C.~C., Bassa, C.~G., Wex, N., et al.\ 2011, 412, 2763
\bibitem[Fruchter et al.(1988)]{1988Natur.333..237F} Fruchter, A.~S., Stinebring, D.~R., \& Taylor, J.~H.\ 1988, 333, 237
\bibitem[Gal-Yam et al.(2006)]{2006Natur.444.1053G} Gal-Yam, A., Fox, D.~B., Price, P.~A., et al.\ 2006, 444, 1053
\bibitem[Gehrels et al.(2006)]{2006Natur.444.1044G} Gehrels, N., Norris, J.~P., Barthelmy, S.~D., et al.\ 2006, 444, 1044
\bibitem[Han et al.(1995)]{1995MNRAS.272..800H} Han, Z., Podsiadlowski, P., \& Eggleton, P.~P.\ 1995, 272, 800
\bibitem[Hansen \& Phinney(1997)]{1997MNRAS.291..569H} Hansen, B.~M.~S., \& Phinney, E.~S.\ 1997, 291, 569
\bibitem[Hartman et al.(1997)]{1997A&A...322..477H} Hartman, J.~W., Bhattacharya, D., Wijers, R., \& Verbunt, F.\ 1997, 322, 477
\bibitem[Hjellming \& Webbink(1987)]{1987ApJ...318..794H} Hjellming, M.~S., \& Webbink, R.~F.\ 1987, 318, 794
\bibitem[Hobbs et al.(2005)]{2005MNRAS.360..974H} Hobbs, G., Lorimer, D.~R., Lyne, A.~G., \& Kramer, M.\ 2005, 360, 974
\bibitem[Hurley et al.(2000)]{2000MNRAS.315..543H} Hurley, J.~R., Pols, O.~R., \& Tout, C.~A.\ 2000, 315, 543
\bibitem[Hurley et al.(2002)]{2002MNRAS.329..897H} Hurley, J.~R., Tout, C.~A., \& Pols, O.~R.\ 2002, 329, 897
\bibitem[Illarionov \& Sunyaev(1975)]{1975A&A....39..185I} Illarionov, A.~F., \& Sunyaev, R.~A.\ 1975, 39, 185
\bibitem[Kasliwal et al.(2012)]{2012ApJ...755..161K} Kasliwal, M.~M., Kulkarni, S.~R., Gal-Yam, A., et al.\ 2012, 755, 161
\bibitem[Katz(1975)]{1975Natur.253..698K} Katz, J.~I.\ 1975, 253, 698
\bibitem[Kennicutt(1998)]{1998ARA&A..36..189K} Kennicutt, R.~C., Jr.\ 1998, 36, 189
\bibitem[Kiel \& Hurley(2006)]{2006MNRAS.369.1152K} Kiel, P.~D., \& Hurley, J.~R.\ 2006, 369, 1152
\bibitem[King et al.(2007)]{2007MNRAS.374L..34K} King, A., Olsson, E., \& Davies, M.~B.\ 2007, 374, L34
\bibitem[Kluzniak et al.(1988)]{1988Natur.334..225K} Kluzniak, W., Ruderman, M., Shaham, J., \& Tavani, M.\ 1988, 334, 225
\bibitem[Lattimer(2012)]{2012AIPC.1484..319L} Lattimer, J.~M.\ 2012, American Institute of Physics Conference Series, 1484, 319
\bibitem[Lipunov(1982)]{1982SvA....26...54L} Lipunov, V.~M.\ 1982, 26, 54
\bibitem[Lipunov et al.(1992)]{1992ans..book.....L} Lipunov, V.~M., B{\"o}rner, G., \& Wadhwa, R.~S.\ 1992, Astronomische Nachrichten Supplement, 108
\bibitem[Liu et al.(2007)]{2007A&A...469..807L} Liu, Q.~Z., van Paradijs, J., \& van den Heuvel, E.~P.~J.\ 2007, 469, 807
\bibitem[Lorimer(2008)]{2008LRR....11....8L} Lorimer, D.~R.\ 2008, Living Reviews in Relativity, 11,
\bibitem[Lovelace et al.(1999)]{1999ApJ...514..368L} Lovelace, R.~V.~E., Romanova, M.~M., \& Bisnovatyi-Kogan, G.~S.\ 1999, 514, 368
\bibitem[L{\"u} et al.(2006)]{2006MNRAS.372.1389L} L{\"u}, G., Yungelson, L., \& Han, Z.\ 2006, 372, 1389
\bibitem[L{\"u} et al.(2008)]{2008ApJ...683..990L} L{\"u}, G., Zhu, C., Han, Z., \& Wang, Z.\ 2008, 683, 990
\bibitem[L{\"u} et al.(2013)]{2013ApJ...768..193L} L{\"u}, G., Zhu, C., \& Podsiadlowski, P.\ 2013, 768, 193
\bibitem[L{\"u} et al.(2017)]{2017ApJ...847...62L} L{\"u}, G., Zhu, C., Wang, Z., \& Iminniyaz, H.\ 2017, 847, 62
\bibitem[L{\"u} et al.(2009)]{2009MNRAS.396.1086L} L{\"u}, G., Zhu, C., Wang, Z., \& Wang, N.\ 2009, 396, 1086
\bibitem[L{\"u} et al.(2012)]{2012MNRAS.424.2265L} L{\"u}, G.-L., Zhu, C.-H., Postnov, K.~A., et al.\ 2012, 424, 2265
\bibitem[Manchester et al.(2005)]{2005AJ....129.1993M} Manchester, R.~N., Hobbs, G.~B., Teoh, A., \& Hobbs, M.\ 2005, 129, 1993
\bibitem[Margalit \& Metzger(2016)]{2016MNRAS.461.1154M} Margalit, B., \& Metzger, B.~D.\ 2016, 461, 1154
\bibitem[Metzger(2012)]{2012MNRAS.419..827M} Metzger, B.~D.\ 2012, 419, 827
\bibitem[Miller \& Scalo(1979)]{1979ApJS...41..513M} Miller, G.~E., \& Scalo, J.~M.\ 1979, 41, 513
\bibitem[Nelemans \& Jonker(2010)]{2010NewAR..54...87N} Nelemans, G., \& Jonker, P.~G.\ 2010, 54, 87
\bibitem[Nelemans et al.(2001)]{2001A&A...375..890N} Nelemans, G., Yungelson, L.~R., \& Portegies Zwart, S.~F.\ 2001, 375, 890
\bibitem[Ohlmann et al.(2016)]{2016MNRAS.462L.121O} Ohlmann, S.~T., R{\"o}pke, F.~K., Pakmor, R., Springel, V., \& M{\"u}ller, E.\ 2016, 462, L121
\bibitem[Paczynski(1986)]{1986ApJ...308L..43P} Paczynski, B.\ 1986, 308, L43
\bibitem[Pan et al.(2013)]{2013Ap&SS.346..119P} Pan, Y.~Y., Wang, N., \& Zhang, C.~M.\ 2013, 346, 119
\bibitem[Paschalidis et al.(2009)]{2009PhRvD..80b4006P} Paschalidis, V., MacLeod, M., Baumgarte, T.~W., \& Shapiro, S.~L.\ 2009, 80, 024006
\bibitem[Perets et al.(2010)]{2010Natur.465..322P} Perets, H.~B., Gal-Yam, A., Mazzali, P.~A., et al.\ 2010, 465, 322
\bibitem[Pfahl et al.(2002)]{2002ApJ...573..283P} Pfahl, E., Rappaport, S., \& Podsiadlowski, P.\ 2002, 573, 283
\bibitem[Piran(1992)]{1992ApJ...389L..45P} Piran, T.\ 1992, 389, L45
\bibitem[Portegies Zwart et al.(2011)]{2011ApJ...734...55P} Portegies Zwart, S., van den Heuvel, E.~P.~J., van Leeuwen, J., \& Nelemans, G.\ 2011, 734, 55
\bibitem[Portegies Zwart \& Yungelson(1999)]{1999MNRAS.309...26P} Portegies Zwart, S.~F., \& Yungelson, L.~R.\ 1999, 309, 26
\bibitem[Pringle \& Rees(1972)]{1972A&A....21....1P} Pringle, J.~E., \& Rees, M.~J.\ 1972, 21, 1
\bibitem[Ruffini et al.(2016)]{2016ApJ...832..136R} Ruffini, R., Rueda, J.~A., Muccino, M., et al.\ 2016, 832, 136
\bibitem[Ryba \& Taylor(1991)]{1991ApJ...380..557R} Ryba, M.~F., \& Taylor, J.~H.\ 1991, 380, 557
\bibitem[Soberman et al.(1997)]{1997A&A...327..620S} Soberman, G.~E., Phinney, E.~S., \& van den Heuvel, E.~P.~J.\ 1997, 327, 620
\bibitem[Stappers et al.(1998)]{1998ApJ...499L.183S} Stappers, B.~W., Bailes, M., Manchester, R.~N., Sandhu, J.~S., \& Toscano, M.\ 1998, 499, L183
\bibitem[Strolger et al.(2004)]{2004ApJ...613..200S} Strolger, L.-G., Riess, A.~G., Dahlen, T., et al.\ 2004, 613, 200
\bibitem[Tauris \& Savonije(1999)]{1999A&A...350..928T} Tauris, T.~M., \& Savonije, G.~J.\ 1999, 350, 928
\bibitem[Thompson et al.(2009)]{2009arXiv0912.0009T} Thompson, T.~A., Kistler, M.~D., \& Stanek, K.~Z.\ 2009, arXiv:0912.0009
\bibitem[Urpin \& Konenkov(1997)]{1997MNRAS.284..741U} Urpin, V., \& Konenkov, D.\ 1997, 284, 741
\bibitem[van den Heuvel \& Bonsema(1984)]{1984A&A...139L..16V} van den Heuvel, E.~P.~J., \& Bonsema, P.~T.~J.\ 1984, 139, L16
\bibitem[van Haaften et al.(2012)]{2012A&A...537A.104V} van Haaften, L.~M., Nelemans, G., Voss, R., Wood, M.~A., \& Kuijpers, J.\ 2012, 537, A104
\bibitem[Yisikandeer et al.(2016)]{2016JApA...37...22Y} Yisikandeer, A., Zhu, C., Wang, Z., \& L{\"u}, G.\ 2016, Journal of Astrophysics and Astronomy, 37, 22
\bibitem[Yungelson(2008)]{2008AstL...34..620Y} Yungelson, L.~R.\ 2008, Astronomy Letters, 34, 620
\bibitem[Yungelson et al.(1993)]{1993ApJ...418..794Y} Yungelson, L.~R., Tutukov, A.~V., \& Livio, M.\ 1993, 418, 794









\end{thebibliography}
\end{document}